\documentclass[sigconf,10pt]{acmart}

\usepackage[T1]{fontenc}
\usepackage{mdwlist}
\usepackage{amsfonts}
\usepackage{graphicx}
\usepackage{xcolor}
\usepackage{url}
\usepackage{xspace}
\usepackage{subfigure}
\usepackage{bbm}
\usepackage{enumitem}

\newcounter{glossy_enum}
\makecompactlist{list*}{list}
\newenvironment{glossy_enumerate}
{\begin{list*}[{\arabic{glossy_enum})}{\usecounter{glossy_enum} \topsep=0.2em \leftmargin=1.4em \itemindent=-0.0em}]}
{\end{list*}\vspace*{0.5em}}

\setenumerate{topsep=0.2em,leftmargin=2.4em,itemindent=-0.0em}

\newcommand{\revision}[1]{\textcolor{black}{#1}}

\newcommand{\fullSVM}{\textsc{Regular SVM}\xspace}
\newcommand{\approxSVM}{\textsc{ASVM}\xspace}
\newcommand{\knn}{\textsc{k-NN}\xspace}
\newcommand{\forest}{\textsc{Forest}\xspace}

\newcommand{\fakepar}[1]{\vspace{0mm}\noindent\textbf{#1.}}
\newcommand\figref[1]{Fig.\,\ref{#1}}
\newcommand\secref[1]{Sec.\,\ref{#1}}

\newcommand{\capt}[1]{}
\newcommand{\code}[1]{\texttt{\textbf{#1}}}

\author[Luca Mottola, Arlsan Hameed, and Thiemo Voigt]{Luca Mottola$^{+*\dagger}$, Arlsan Hameed$^{\dagger}$, and Thiemo Voigt$^{*\dagger}$\\
$^{+}$Politecnico di Milano (Italy), $^{*}$RI.SE, $^{\dagger}$Uppsala University, Sweden}

\begin{document}

\title{Uncharted Territory: \\Energy Attacks in the Battery-less Internet of Things}

\begin{abstract}

We study how ambient energy harvesting may be used as an attack vector in the battery-less Internet of Things~(IoT).
\revision{Battery-less IoT devices rely on ambient energy harvesting and are employed in a multitude of applications, including safety-critical ones such as biomedical implants.}
Due to scarce energy intakes and limited energy buffers, their executions become \emph{intermittent}, alternating periods of active operation with periods of recharging energy buffers.
Through an independent exploratory study and a follow-up systematic analysis, we demonstrate that by exerting limited control on ambient energy one can create situations of \emph{livelock}, \emph{denial of service}, and \emph{priority inversion}, without physical device access.
We call these situations \emph{energy attacks}.
Using concepts of approximate intermittent computing and machine learning, we design a technique that can detect energy attacks with 92\%+ accuracy, that is, up to 37\% better than the baselines, and with up to one fifth of their energy overhead.
Crucially, \emph{by design}, our technique does \emph{not cause any additional energy failure} compared to the regular intermittent processing.
We conclude with directions to inspire defense techniques and a discussion on the feasibility of energy attacks.

\end{abstract}


\maketitle              

\section{Introduction}

\revision{Ambient energy harvesting allows Internet of Things~(IoT) devices to eliminate their dependency on traditional batteries~\cite{harvesting-survey}.
This enables drastic reductions of maintenance costs and previously unattainable deployments, even in safety-critical settings such as biomedical implants~\cite{water-deployment-microbial-fuel-cell,tethys,soil-termoelectric,sensys20deployment,denby2023kodan, harvesting-survey}.}

\revision{Harvested energy is generally highly variable in time~\cite{harvesting-survey}, yet energy buffers, such as capacitors, need to be miniaturized as well to limit device footprint, and therefore offer limited energy budgets.}
System shutdowns due to energy depletion are unavoidable and computing becomes \emph{intermittent}~\cite{awesome}: periods of active execution and periods of energy harvesting come to be unpredictably interleaved.

\begin{figure}[tb]
  \centering
    \includegraphics[width=.93\linewidth]{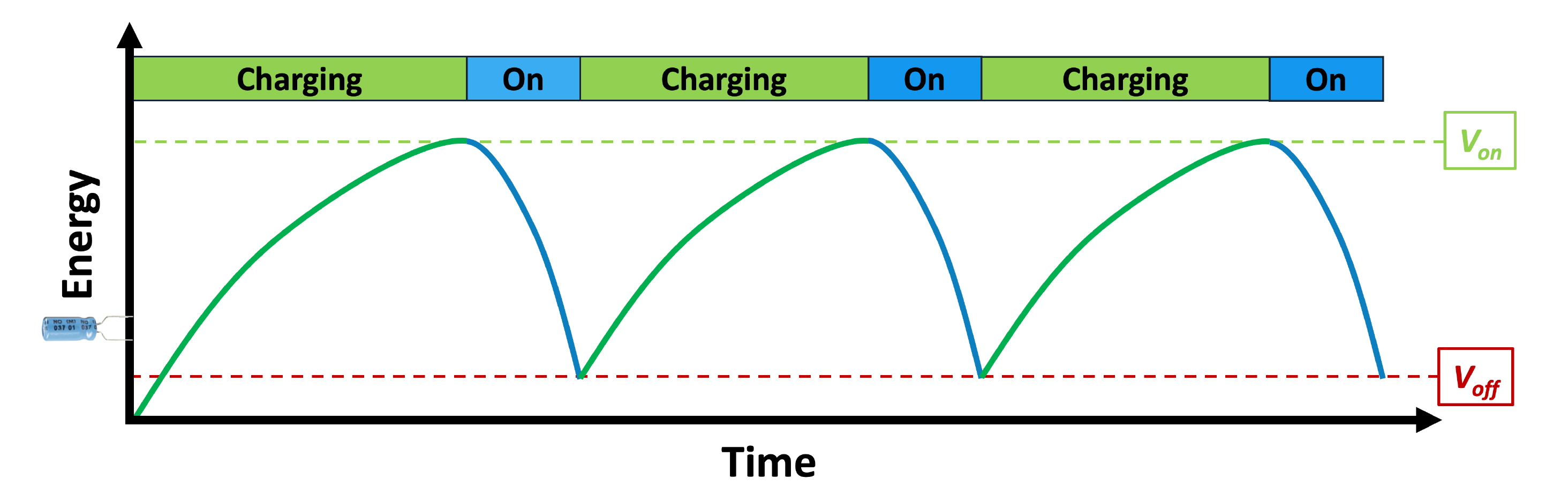}
    \vspace{-2mm}
   \caption{Example intermittent execution. \capt{Periods of active computation and periods of charging the energy buffer alternate.}}
    \vspace{-4mm}
   \label{fig:execution}
 \end{figure}

\fakepar{Computing intermittently} \figref{fig:execution} shows an example execution.
The ambient charges the onboard capacitor until voltage~$V_{\mathit{on}}$ is reached that causes the device to power on.
The device senses, computes, and communicates as long as the capacitor charge remains above a threshold $V_{\mathit{off}}$.
The device then switches off, waiting for the capacitor to reach $V_{\mathit{on}}$ again.
This pattern may occur on tiny time scales; computing simple error correction codes on a battery-less IoT device may require as many as 16 energy cycles~\cite{HarvOS}.

Due to resource constraints, applications run with \emph{no operating system support}~\cite{awesome}.
When the device powers off at~$V_{\mathit{off}}$, the system state would normally be lost.
Intermittently-computing IoT systems use checkpointing~\cite{HarvOS,mementos,Hibernus,Hibernus++,chinchilla,QuickRecall,Samoyed,ratchet,dice} or task-based programming~\cite{alpaca,chain,DINO,Coati,ink,coala} to create persistent state on non-volatile memory (NVM).
Operations on NVM, however, are extremely energy hungry~\cite{maioli2021alfred}.

 \fakepar{Energy harvesting as attack vector} This paper is about a \emph{new, largely unexplored direction} in intermittent computing: we study how exerting limited control on ambient energy supplies may steer intermittent executions in unintended ways.
The simplest scenario consists, for example, in physically blocking the solar radiation arriving at a solar panel that powers the device, eventually impeding forward progress.
We demonstrate that much more subtle situations exist.
We generally call these situations \emph{energy attacks}.

\revision{We start from an independent exploratory study to understand whether vulnerabilities exists that may be exploited through energy attacks, and to check their practical feasibility.
 The results, reported in \secref{sec:reality}, indicate that it takes no more than \emph{a few weeks} for M.Sc.-level students with \emph{no} specific training to identify vulnerabilities and use energy attacks to exploit them.}
These insights represent the motivation for our work, which is centered on two cornerstones. 

First, in \secref{sec:attacks} we define the attack model and systematically analyze the vulnerabilities discovered during the exploratory study, while independently discovering a further vulnerability.
We experimentally demonstrate that energy attacks can exploit these vulnerabilities to create situations of \emph{livelock}, \emph{priority inversion}, and \emph{denial of service}.
Unlike the simplest scenario above, where a permanent energy blockage trivially brings the system to a complete halt, we provide quantitative evidence that these attacks create situations that are deceptively similar to regular energy patterns.

Next, in \secref{sec:detection} we tackle the problem of \emph{detecting} energy attacks.
The problem appears as a case of \emph{anomaly detection}~\cite{chandola2009anomaly}. 
Three peculiar requirements exist: detecting energy attacks \emph{i)}~\emph{accurately} and \emph{ii)} \emph{with low latency}, while doing so \emph{iii)} \emph{right on the IoT devices}, as opposed to an external system, to spare the energy overhead of radio operations.
Our detection technique uses concepts of approximate intermittent computing~\cite{bambusi2022case,umesh2021survey} and machine learning to ensure that, \emph{by design}, the attack detection process \emph{does not cause additional energy failures} compared to the regular intermittent processing, and hence bears minimal impact.

\secref{sec:eval} reports on the accuracy and overhead of our detection technique, based on 500K+ data points obtained using real-world energy traces, compared with two baselines.
The results indicate that our technique is 92\%+ accurate, which is up to 37\% better than the baselines, and imposes an overhead that is up to one fifth of the baselines.
Further, the detection performance is largely independent of the energy patterns and robust to previously unseen attacks.

Following detection of an energy attack, the system should apply countermeasures to limit the negative effects. 
In \secref{sec:defense}, we explore the multiple dimensions of the problem, articulate the related trade-offs, and offer arguments to inspire the design of defense techniques.
The latter include attack-specific countermeasures as well as generic software and/or hardware techniques, such as adaptive energy management and supporting multiple energy sources~\cite{liu2021hybrid}.
We end the paper with a discussion of our work's limitations and of the feasibility of energy attacks in \secref{sec:discussion}.

Before moving to the subject matter, we illustrate necessary background information and survey related work . 



\section{Background and Related Work}
\label{sec:related}


We provide background information and survey related works from data centers to battery-powered and battery-less IoT.

\fakepar{Power attacks in data centers} Our work resembles some similarities with power attacks in data centers.
With the increasing number of physical servers, their power distribution systems tend to approach peak capacities and power oversubscription is used to handle power provisioning.
This works as long as servers do not peak simultaneously.
However, malicious workloads may generate power spikes on multiple servers at the same time, which causes branch circuit breakers to trip, leading to power outages~\cite{greenberg2008cost,li2016power}.
Virtual machine provisioning~\cite{chen2020power} and side channels reporting thermal information of servers~\cite{chen2020power} are used to create abnormal behaviors.
Detection techniques include machine learning applied to performance logs~\cite{chen2020power} and modeling user behaviors that may indicate the infrastructure is under-performing~\cite{li2016power}.

Common with our problem is that energy is part of the attack vector.
However, the technology is extremely different, for example, in terms of workloads and hardware platforms.
Moreover, in contrast to the attack model we describe in \secref{sec:attacks}, attackers do not directly manipulate the energy provisioning channel and need access to the target data center or must be informed of its layout.

\fakepar{Security in battery-powered IoT} Resource-constrain\-ed IoT devices are difficult to secure due to resource constraints, which complicates the use of regular security mechanisms~\cite{thakor2021lightweight}.

Battery-powered IoT devices enables peculiar attacks, for example, in an attempt to drain batteries~\cite{krentz2017countering,nguyen2019energy}.
Low-power radios make IoT devices vulnerable to denial of service attacks, for example, due to intentional jamming~\cite{kanwar2021jamsense}.
Multi-hop networks require specialized network stacks that open to new kinds of attacks, in particular at the routing layer, motivating new security mechanisms ranging from hardware-based solutions~\cite{portilla2010adaptable} to methods for attack detection and mitigation that rely on machine learning~\cite{da2019internet,tahsien2020machine}.

\revision{These approaches, unfortunately, falls short of expectations for battery-less IoT devices, where energy constraints are way more severe.
  Moreover, intermittent executions add a new dimension to the problem that requires specialized solutions, as we explain next.}

\fakepar{Intermittent computing} The prevailing architecture includes a mixed-volatile MCU~\cite{msp430fr5969} with built-in NVM for persisting state, and a capacitor  to tame fluctuations of energy intake.
Such device configuration is seen in both available platforms~\cite{capacity-over-capacitance,flicker} and concrete deployments~\cite{water-deployment-microbial-fuel-cell,pible}.


Specialized architectures also exist that use separate capacitors of different sizes as energy buffers~\cite{capybara}.
This allows the system to strike a better trade-off between charging times and available energy.
Smaller capacitors are the first to reach $V_{on}$; as this happens, tasks that consume little energy, such as probing low-power sensors, are immediately executed.
Bigger capacitors take longer to reach $V_{on}$; their energy is eventually consumed by energy-hungry tasks, such as controlling actuators or radio operations.

Checkpointing~\cite{HarvOS,mementos,Hibernus,Hibernus++,chinchilla,QuickRecall,Samoyed,ratchet,dice} or task-based programming~\cite{chain,DINO,Coati,ink,coala} is used to deal with energy failures.
The former consist in replicating the application state on NVM, where it is retrieved back once the system resumes with sufficient energy.
The latter offer abstractions to define and manage persistent state, while taking care of data consistency in case of repeated executions of non-idempotent code~\cite{ratchet}.
Software techniques are also used to handle peripheral states across energy failures~\cite{karma} and estimate energy consumption~\cite{epic}.


\fakepar{Security and intermittent computing} The security scenario becomes uncharted territory here.

The few existing solutions focus on securing persistent state.
Krishnan et al.~\cite{Krishnan2018} demonstrate that persistent state is vulnerable to sniffing, spoofing, or replay attacks.
Asad et al.~\cite{asad2020securing} experimentally evaluate the use of different encryption algorithms and ARM TrustZone protection.
Krishnan et al.~\cite{10.1145/3522748} build on this and propose a configurable checkpoint security setting that leverages application properties to reduce overhead. 
Ghodsi et al.~\cite{ghodsi2017optimal} use lightweight algorithms~\cite{borghoff2012prince} for securing checkpoints, ensuring confidentiality.
Valea et al.~\cite{valea2018si} propose a SECure Context Saving 
hardware module inside the MCU.
In contrast, Grisafi et al.~\cite{grisafi2022mpi} present a hypervisor to manage and protect checkpoints.
Khrishnan et al.~\cite{krishnan2019secure} present a generic secure protocol and apply Authenticated Encryption with Associated Data 
to protect checkpointing data.

\revision{Unlike the works above, we study new types of attacks realized by exerting control on ambient energy provisioning.
The original motivation for our work is the outcome of an independent exploratory study we run, illustrated next.}


\section{Motivation}
\label{sec:reality}

Battery-less IoT deployments appear in ever increasing and diverse scenarios, including safety-critical ones~\cite{water-deployment-microbial-fuel-cell,tethys,soil-termoelectric,sensys20deployment,denby2023kodan}.
Securing their operation is crucial.
To understand the current state of affairs, we run an \emph{independent exploratory study}.

\fakepar{Setting} We recruit eight computer engineering M.Sc. students\footnote{We obtained IRB approval from their institution. They are not compensated.}.
They attended courses in low-power wireless networks, embedded programming, and IoT.
They have \emph{no} earlier training on battery-less IoT devices.
We hand them reading material to gain knowledge of existing techniques~\cite{awesome,HarvOS, alpaca}, and provide each of them with a TI MSP430FR5969 Launchpad attached to either a 40 x 40 mm polycrystalline silicon solar panel or to a Powercast receiver, plus a two-capacitor energy subsystem~\cite{capybara} that we emulate using an Arduino Uno board as energy controller.
Each Powercast receiver is paired to a single Powercast transmitter.
Each students can choose which energy harvester to use.

We give the students the C implementation of a typical sense-process-transmit loop~\cite{water-deployment-microbial-fuel-cell,tethys,soil-termoelectric,sensys20deployment}.
After a single day of training, the students can run the code on the Launchpads.
Six students choose to use HarvOS, an existing checkpointing system for intermittent computing~\cite{HarvOS} and only minimally refactor the original application code.
Out of the six students, four use the solar panel and the other two use the Powercast system.
The remaining two students use the solar panel and and split the code in tasks, using Alpaca~\cite{alpaca}.

We challenge the students with the following goal: \emph{without changing the code, disrupt the application without completely halting the system or physically accessing the device.}
The students work independently and are not allowed to exchange ideas, information, or code.
We also make sure they cannot rely on backchannel information from each other, for example, by sniffing wireless transmissions.
Besides observing their work, we conduct semi-structured interviews at the end of every day.
The transcripts are available~\cite{interviews}.

\fakepar{Outcome} The students recognize that, without physical access to the device, \emph{energy harvesting is a potential attack vector}.
Permanently blocking the energy source, however, would bring the system to a complete halt.
 
After two full days of experimentation, five of the six students using HarvOS \emph{figure out they can break the system by creating a livelock}.
During the interview, four students explicitly mention``livelock'', while the fifth student intuitively describes what is, in fact, a livelock.
They spend the following days trying to create the conditions that lead to the livelock.
Depending on the harvesting technology, they end up conceiving different techniques to achieve this.

Two students using the solar panel succeed halfway through the third day by \emph{intentionally} alternating periods of blocking the energy source with periods of regular operation, based on information from sniffed packets indicating at what stage is the execution.
The third student using the solar panel succeeds using a similar technique half a day later.
After two weeks, one student succeeds in making the attack automatic.
Using a light sensor on a separate device, he estimates the amount of energy harvested by the target device and accordingly tunes the periods of energy blockage. 

After about five days of work, the other two students using HarvOS and the Powercast system create a setup with a second Powercast receiver-transmitter pair generating a signal \emph{in phase opposition} with the regular one, yielding destructive interference.
To identify the ``right'' phase, they attach the second receiver to a laptop that controls the second Powercast transmitter.
The laptop replicates the regular signal obtained from the attached Powercast receiver with a variable phase within a $[-\lambda/2, \lambda/2]$ interval, at small increments of $\delta$.
Sniffing packets from the target device allows them to determine when to stop the process, which corresponds to when they hit ``right'' phase that almost cancels out the original energy signal~\cite{liu2016safe,naderi2014rf}.
By doing so periodically, they eventually achieve the same effect as with the solar panel.

It takes a bit more for the two students using solar panels and Alpaca to \emph{realize they can fiddle with the multiple capacitors}.
After about two weeks, each of them creates a setup that blocks the energy arriving at the largest capacitor to slow down the task that takes energy from that.
This, combined with the larger leakage compared to the smaller capacitor, makes the task powered by the larger capacitor progressively diminish the execution frequency.
Because the tasks powered by the smaller capacitors continue producing data almost at the regular rate, without the tasks powered by the larger capacitors consuming it at comparable pace, the buffers between the tasks eventually overflow.

\fakepar{Conclusion} In a matter of weeks, seven computer engineering M.Sc. students out of eight, with no previous exposure to the technology at stake, managed to successfully setup energy attacks.
The striking conclusion we draw is that the technology is \emph{vulnerable}.

Likely because research efforts in intermittent computing concentrate on issues such as maintaining forward progress, understanding the related security issues is much less investigated.    
This includes both investigating the effects of known attacks and \emph{exploring new types of threats}.


\section{Energy Attacks}
\label{sec:attacks}

We systematically analyze the vulnerabilities discovered in the exploratory study and investigate a further vulnerability we discover independently.
To this end, we describe first the attack model we adopt; next, we show evidence of vulnerabilities and scrutinize the corresponding energy attacks. 

\subsection{Attack Model}

\begin{figure}[tb]
  \centering
    \includegraphics[width=.99\linewidth]{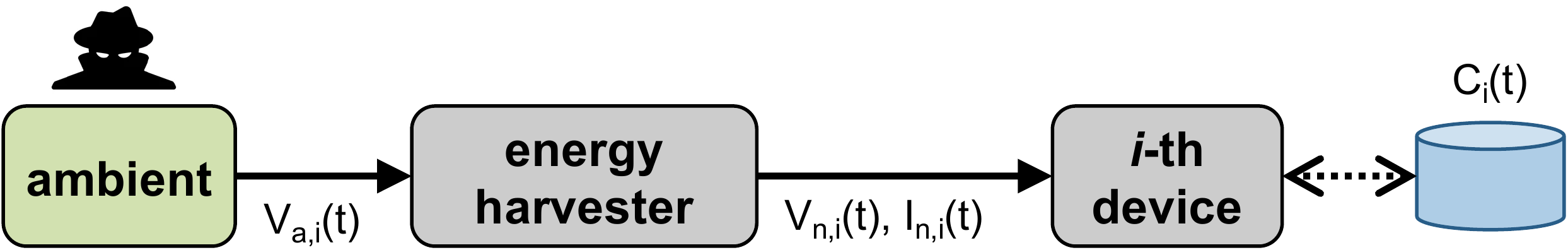}
    \vspace{-2mm}
   \caption{Attack model. \capt{The attacker manipulates the source or intervenes along the path from the source to the harvester.}}
    \vspace{-2mm}
   \label{fig:attackmodel}
\end{figure}

\figref{fig:attackmodel} illustrates the attack model we adopt.
A resource-constrained intermittently-computing IoT device is equipped with multiple sensors and/or actuators, an MCU, a radio, and an energy management circuitry attached to the output of the energy harvester and used to charge the local energy buffer.
Such configuration is seen in multiple real-world deployments of battery-less IoT devices~\cite{water-deployment-microbial-fuel-cell,tethys,soil-termoelectric,sensys20deployment}.

The energy coming from the ambient $a$ and arriving at the energy harvester of node $i$ is modeled as a continuous signal of voltage~$V_{a,i}(t)$.
This describes the energy content made available by the ambient to node $i$ at time~$t$.
\revision{For simplicity, our description here considers a single energy source.
The corresponding analysis, however, applies no matter the number of energy sources, as long as the attack model is applicable to each of them.
Relying on multiple energy sources may be, nonetheless, a way to defend against energy attacks.
We discuss this aspect further in \secref{sec:defense}.}

The energy harvester of node $i$ takes $V_{a,i}(t)$ as input and transforms it into an energy signal of voltage $V_{n,i}(t)$ and current $I_{n,i}(t)$.
The latter is a function of $V_{n,i}(t)$ and of the equivalent resistance offered by the charging circuitry at node~$i$.
The energy signal described by $V_{n,i}(t)$ and $I_{n,i}(t)$ charges the local energy buffer, eventually discharged while sensing, computing, or communicating.
We model the charge available in the energy buffer of node $i$ as $C_i(t)$.

The attacker has no physical access to the devices and no knowledge of the relation between $V_{a,i}(t)$ and $V_{n,i}(t)$ or $I_{n,i}(t)$.  
She can, however, sniff packets, inspect their content, 
and \revision{intervene along the path from the energy source to the energy harvester attached to the device, including directly controlling the energy source}.
\revision{This means the attacker can alter the value of $V_{a,i}(t)$ taken as input by the energy management circuitry at node~$i$.}
We model this as a function $a_i(V_{a,i}(t))$, that is, a transformation $a$ from the voltage domain to the same domain, specific to node $i$.

An elementary example of function $a$ that causes a denial of service at node $i$ from $t'$ onwards is $a_i(V_{a,i}(t)) = 0,t>t'$, that is, the harvester at node $i$ receives no energy after $t'$.
The energy buffer at node $i$ progressively discharges because of application processing and capacitor leakage, until the device persists the state before entering the charging phase, as shown in \figref{fig:execution}.
However, because $a_i(V_{a,i}(t)) = 0, t>t'$, that is, there is no energy arriving at node $i$ later than $t'$, $C_i(t)$ never reaches $V_{on}$ again, and node $i$ never resumes.

\revision{The application's source code is available or can be reverse-engineenered from the binaries~\cite{udupa2005deobfuscation}. 
Codebases for battery-less IoT systems, including and especially the ones used in real deployments~\cite{water-deployment-microbial-fuel-cell,sensys20deployment,denby2023kodan}, are often public~\cite{amiri2017survey}, including operating system layers~\cite{awesome} and hardware drivers~\cite{karma,sytare,restop}. 
Compilers~\cite{gay2003nesc} often require the entire source code to perform full-program optimizations.}

\subsection{Evidence of Vulnerabilities}

We systematically analyze and experimentally demonstrate the two vulnerabilities discovered during the exploratory study. 
\revision{Both target single devices; their individual placement in space with respect to each is therefore immaterial.
We also study a further vulnerability we discover independently, which affects a network of battery-less IoT devices.}
Energy attacks exploiting these vulnerabilities may lead to \emph{livelocks}, \emph{denial of service}, or \emph{priority inversion}.

Unless otherwise specified, we consider a TI MSP430FR5969 running at 1 MHz as target MCU and set $V_{\mathit{on}}$ to 3.3 V and $V_{\mathit{off}}$ to 1.8~V, which is the most energy-efficient setting~\cite{ahmed2020intermittent}.
We use different methodologies and tools, including real hardware, numerical simulations, and time-accurate emulation.

\fakepar{Attack \#1: static activation thresholds $\Rightarrow$ livelock} Multiple students in the exploratory study find out that a vulnerability exists in how system software for intermittent computing is often designed.
That allows an attacker to create a situation of livelock, that is, a condition where the system keeps repeating the same set of operations, and yet makes no progress in the long run~\cite{cleancut}.

Established design processes for intermittent systems recommend setting the activation threshold $V_{\mathit{on}}$ by striking a balance between charging times and energy content at $V_{\mathit{on}}$~\cite{HarvOS,mementos,Hibernus,QuickRecall,Samoyed,dice,capybara,flicker,awesome}.
The former suggests a lower $V_{\mathit{on}}$, whereas the latter pushes for a higher $V_{on}$. 
In most existing systems~\cite{HarvOS,mementos,Hibernus,QuickRecall,Samoyed,dice}, $V_{\mathit{on}}$ is \emph{statically} set before deployment and does not necessarily guarantee that the energy content is sufficient to make progress in the application logic \emph{and} persist the new state before an imminent energy failure.
The ambient is indeed assumed to provide some energy also \emph{during} the active times, making it possible to partially replenish the energy buffer while the system progresses~\cite{heliomote,awesome}.

An attacker may, however, systematically block the energy source at a node $i$ while the device is computing, that is, he creates a transformation $a$  such that $a_i(V_{a,i}(t)) = 0, t'>t>t''$, where $t'$ and $t''$ are the points in time where the system reaches $V_{\mathit{on}}$ and $V_{\mathit{off}}$, respectively.
We demonstrate this situation using one of the students' prototype of \secref{sec:reality}, including a TI MSP430FR5969 Launchpad and two Powercast transmitter-receiver pairs, running HarvOS~\cite{HarvOS}.

While the first pair of Powercast transmitter-receiver normally powers the device, the attacker implements function~$a$ by controlling when to generate the opposing signal with the second Powercast transmitter.
Phase alignment is obtained using the same procedure used by the student, as in \secref{sec:reality}.
The attacker identifies $t'$ and $t''$ by correlating wireless traffic to different points in the application execution.

\begin{figure}[tb]
  \centering
    \includegraphics[width=.9\linewidth]{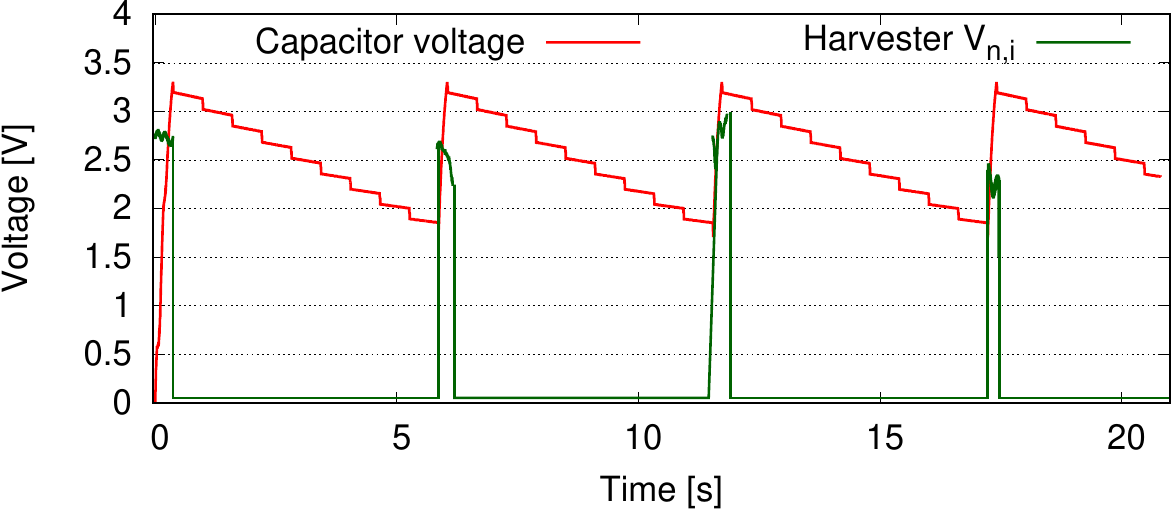}
    \vspace{-2mm}
   \caption{Livelock situation caused by systematically blocking the energy source while the device is active. \capt{The discharge pattern is identical every time the device is computing: it always resumes from the same checkpoint and repeats the same set of instructions, without ever making progress since there is no energy left to persist the state when reaching $V_{\mathit{off}}$.}}
   \label{fig:attack}
    \vspace{-2mm}
 \end{figure}

\figref{fig:attack} shows an example execution once the phase alignment is achieved.
Without any contribution of energy during the active times and a $V_{\mathit{on}}$ setting that does not account for this, the system approaches $V_{\mathit{off}}$ with insufficient energy to persist state, that is, no new checkpoint is created.
When the system reaches $V_{\mathit{on}}$ again, HarvOS resorts to the previous checkpoint, that is, the one that does not include the progress achieved between  $t'$ and $t''$.
The previous operations are then executed again, and with the attacker replaying the same function $a$ once more, the system approaches $V_{\mathit{off}}$ again with insufficient energy to persist the state.
As long as the attacker keeps doing so, the system continues to restart from the same checkpoint, each time executing the same operations, and yet making no progress in the long run.

Situations when the ambient provides no energy during executions may happen, yet not systematically~\cite{mementos,cleancut}.
This is a hint to defend against this attack, as discussed in \secref{sec:defense}.

\fakepar{Attack \#2: skewed energy management $\Rightarrow$ priority inversion} The students using Alpaca discover that multi-capacitor architectures, described in \secref{sec:related}, are vulnerable to an attack that creates a situation akin to priority inversion\footnote{We admit to abuse the term ``priority inversion'' somehow. The system does not run a priority-based scheduler. The priority here is implicitly dictated by the need to eventually run a consumer task before data in the incoming buffer overflows; this attack prevents that from happening.}.

An attacker may ration the energy arriving at the harvester so that the smaller capacitors reach $V_{on}$ more often than the bigger ones, while relying on the larger leakage of the latter to further slow down their charging.
This makes producer tasks, such as probing low-power sensors, push data in the local data buffers more rapidly than tasks such as radio transmissions, which consume the data.
The local buffers eventually overflow.
The students use simple energy profiling tools~\cite{dice} to tune this attack.
Note that from an outsider perspective, the loss of data due to this attack is indistinguishable from other data losses, for example, due to packet losses during wireless transmissions.

\begin{figure}[tb]
  \centering
    \includegraphics[width=.9\linewidth]{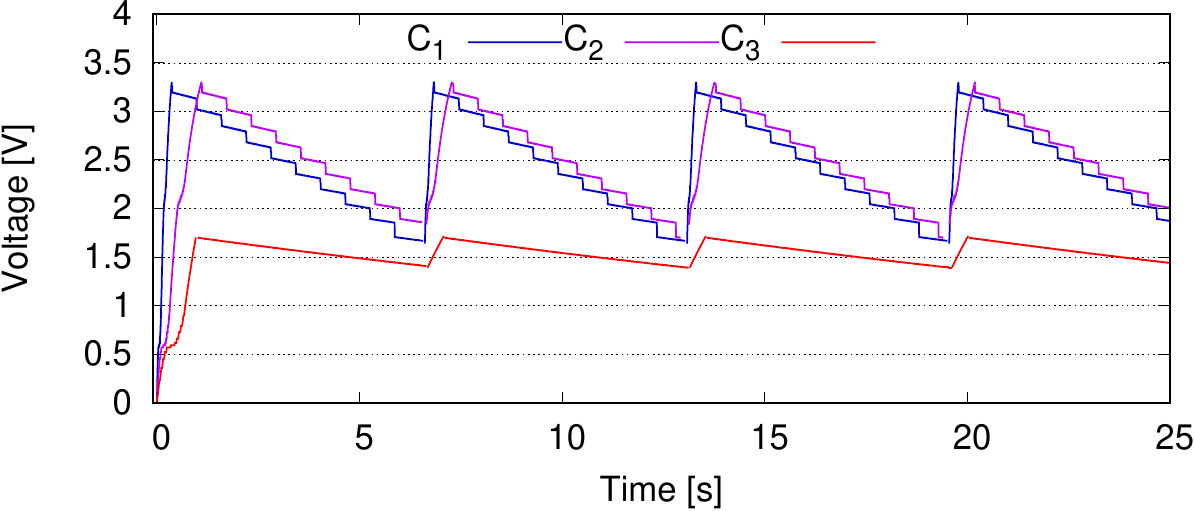}
    \vspace{-2mm}
   \caption{A case of priority inversion generated by purposely suspending energy provisioning. \capt{The largest capacitor $C_3$ is never fully charged and the associated task never executes. This task stops consuming data from a non-volatile queue that eventually overflows.}}
    \vspace{-2mm}
   \label{fig:alpaca}
\end{figure}

To gain a precise understanding of the execution, we investigate this attack using a custom version of the Siren MSP430 emulator~\cite{siren} we develop, which implements the same multi-capacitor hardware architecture~\cite{capybara} used by the students and can re-play existing energy traces from an RF energy source~\cite{mementos}.
The entire codebase of the simulator and the data enabling the study that follows are available~\cite{priorityattack}.
We use three different capacitors $C_1$, $C_2$, and $C_3$, with $C_3 > C_2 >C_1$.
We use $C_1$ for sensing from a low-power temperature sensor, $C_2$ to locally process the data, and $C_3$ for radio operation.

\figref{fig:alpaca} shows the voltage levels at the three capacitors in an example execution.
By periodically interrupting energy provisioning for sufficiently long that $C_3$ is never fully charged, the local data buffer eventually overflows.
This situation is surprisingly easy to reach: non-volatile memories are extremely limited in size; therefore, the local data buffers are normally dimensioned to store just a handful of entries and it does not take long until they fill up.

\fakepar{Attack \#3: unwanted synchronization $\Rightarrow$ denial of service} We investigate ourselves whether vulnerabilities exist in the design of battery-less IoT systems that impact the network as opposed to single devices.
We eventually figure out that energy attacks may be setup so that the wireless transmissions of two or more devices systematically collide, degrading system performance.

\revision{As it is common in many IoT applications~\cite{water-deployment-microbial-fuel-cell,tethys,soil-termoelectric,sensys20deployment}, homogeneous devices are deployed in large numbers with identical hardware and perform the same sense-process-transmit loop~\cite{water-deployment-microbial-fuel-cell,tethys,soil-termoelectric,sensys20deployment}.
Co-located devices are therefore subject to almost identical energy patterns from the ambient~\cite{harvesting-survey, islam2023amalgamated,majid2020continuous}.
An example is when relying on light sources~\cite{islam2023amalgamated,majid2020continuous}.
In these settings, $V_{a,i}(t)$ is the same for all $i$.}

The simplest way of performing this attack is first to totally block the energy intake for the nodes under attack, that is, $a_i(V_{a,i}(t)) = 0, \forall t > t'$.
Eventually, $C_i(t''')=0$ at some point $t'''>t'$ for all nodes $i$ under attack, because even if the devices do not compute, capacitors experience leakage current and self-discharge.
At this point the attacker removes the blockage, restoring the original energy intake with \mbox{$a_i(V_{a,i}(t)) = \mathbbm{1}$}, $\forall t > t'''$.
All capacitors start charging in the same way and devices start operating simultaneously when they reach $V_{on}$.
At this point, they are synchronized.

\begin{figure}[!tb]
\begin{center}
     \includegraphics[width=.90\linewidth]{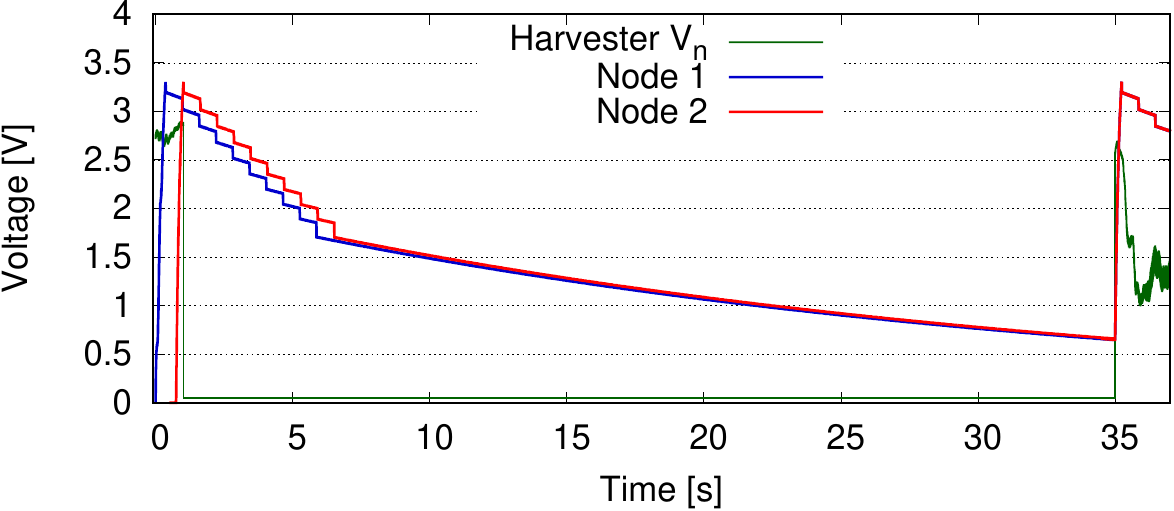}
   \vspace{-2mm}
    \caption{Denial of service created by artificially synchronizing nodes. \capt{By exploiting the natural capacitor leakage, an attacker may synchronize nodes so their packet transmissions systematically collide, causing data losses.}}
\vspace{-2mm}
\label{fig:synctrace}
\end{center}
\end{figure}


To investigate this situation, we develop a custom discrete-event simulator using \code{SimPy}.
The simulator includes accurate numerical models of the essential electronics, including the energy harvester, the energy management circuitry attached to the output of the energy harvester, the capacitor with its leakage, and a load that models the IoT device.
The input to the simulator is a real-world voltage from a solar panel deployed indoor~\cite{epic}.
As for the second attack, the entire codebase of the simulator, along with the data enabling the study that follows, is available~\cite{synchattack}.

\figref{fig:synctrace} 
shows an execution of two devices $1$ and~$2$ where the green curve represents $V_{n,1}(t) = V_{n,2}(t)$, whereas the red and blue curves represent $C_1(t)$ and $C_2(t)$, respectively.
The attacker blocks the energy source at time $t'=1$ s; both $C_1$ and $C_2$ decrease rapidly while the devices continue the execution.
When device~$2$ dumps the state on persistent storage and switches off at time $t''=7$~s, being $a_1(V_{a,1}(t))=a_2(V_{a,2}(t)) = 0, t' < t < t'''$, $t'''=35$~s, and provided the duration of the blockage $\Delta t = t'''-t'$ is sufficiently large, capacitor leakage eventually leads to $C_1(t) \approx C_2(t)\approx 0$.

At time $t'''$, the attacker removes the blockage, therefore $a_1(V_{a,1}(t))=a_2(V_{a,2}(t))=\mathbbm{1}, t> t'''$.
From now on, the two capacitors charge up in the same way, as they are subject to the same ambient energy: the read and blue curves in \figref{fig:synctrace} almost perfectly overlap. 
The two devices reach $V_{on}$ at the same time and restart their execution by going through the same operations at the same times, leading to transmitting simultaneously, generating a packet collision.

Note that the attacker has no information on $C_1$ or $C_2$.
The attack may thus be unsuccessful if $\Delta t$ is too small; for example, because $C_1$ and $C_2$ do not arrive at roughly the same energy content, which is a prerequisite for generating the synchronous execution afterwards.
The attacker may simply play this attack with increasing $\Delta t$ until successful.
Even if the nodes involved are not perfectly synchronized, moving their transmissions closer in time puts increasing pressure on collision avoidance and backoff techniques, which are confronted with an artificial situation that would otherwise be extremely rare.
Network throughput and packet latency are consequently degraded.  

Two key parameters that determine the probability of success of this attack are $\Delta t$ and capacitor size.
Using our simulation tool, we evaluate their impact on the attack's success probability with two nodes.
As expected, the probability of a successful attack increases with~$\Delta t$.
This is because the longer is the time the attacker blocks the energy source, the higher are the chances that eventually $C_1(t) \approx C_2(t)\approx 0$. 
With a smaller capacitor, smaller $\Delta t$ are sufficient for a successful attack, as $C_1(t) \approx C_2(t)\approx 0$ is reached faster.
Further quantitative data is available in an accompanying report~\cite{synchattack}.

Many traits of this scenario are found in real deployments~\cite{water-deployment-microbial-fuel-cell,tethys,soil-termoelectric,sensys20deployment}.
\revision{Relying on the assumption that energy intakes at co-located nodes are similar is \emph{not just common, but even used as a basis to implement communication protocols} in networks of intermittently-computing IoT devices~\cite{geissdoerfer2022learning}.}


\section{Attack Detection}
\label{sec:detection}



Based on the analysis of existing vulnerabilities, we study the problem of \emph{detecting} energy attacks.
This means understanding when ambient energy does not follow the expected patterns, which therefore represents a case of anomaly detection~\cite{chandola2009anomaly}.
Most deployments of IoT battery-less devices operate in areas with little to no opportunities to instrument the area~\cite{water-deployment-microbial-fuel-cell,tethys,soil-termoelectric,sensys20deployment,denby2023kodan}; for example, by installing systems to detect physical intrusion of the attacker using surveillance cameras.
We are, therefore, to meet peculiar requirements:
\begin{description}
\item[R1] \emph{be accurate}, that is, minimizing false positives and false negatives so a device can rely on factual information;
\item[R2] \emph{operate online with low latency}, as countermeasures may be effective only in the short term;
\item[R3] \emph{run locally on the IoT device} due to the excessive energy consumption and additional latency that offloading the process to a third party would incur.
 \end{description}

We articulate next our exploration of the design space and the design of our solution to the problem.

 \subsection{Design Space}

 \begin{figure}[tb]
  \centering
    \includegraphics[width=.9\linewidth]{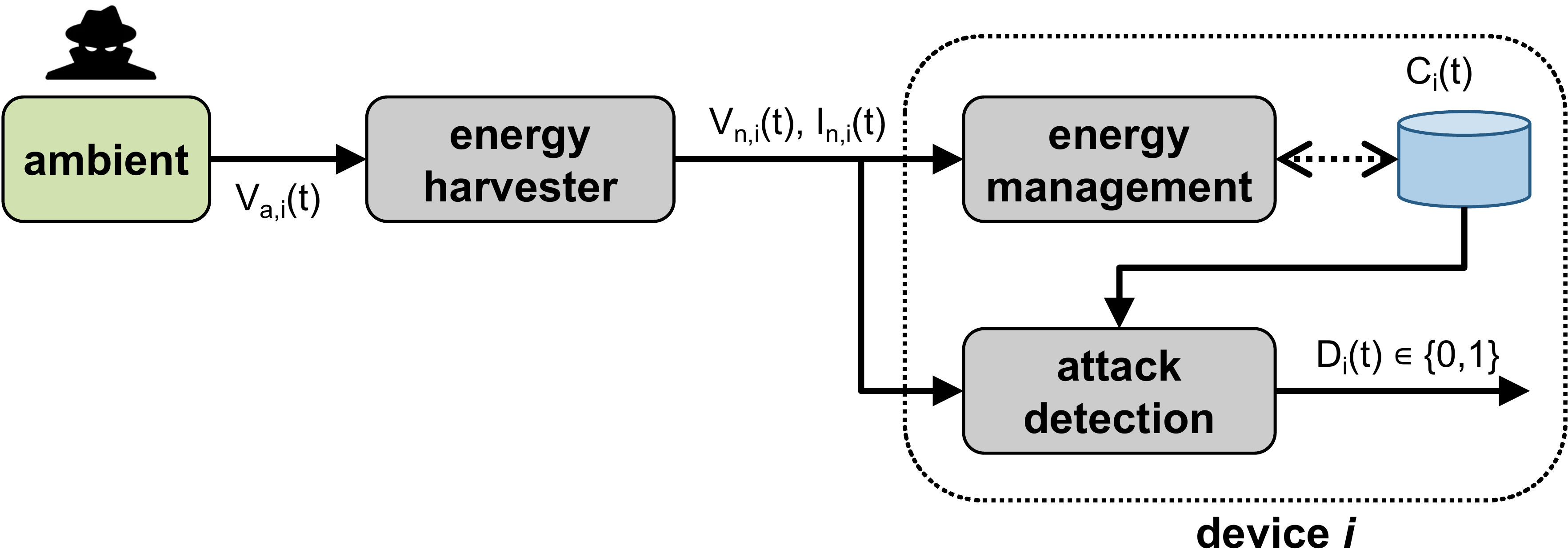}
    \vspace{-2mm}
   \caption{Detection system architecture. \capt{The energy attack detection system can only rely on ambient information after its transformation into usable energy by the harvesting mechanism. Its output is a binary flag $D_i(t) \in \{0,1\}$ indicating whether an attack is ongoing at time $t$.}}
    \vspace{-2mm}
   \label{fig:detection}
 \end{figure}

\figref{fig:detection} shows the architecture of the detection system we are to design.
The detection process runs periodically, every $T$.
We indicate with $D_i(t) \in \{0,1\}$ a binary flag that indicates whether the attack detection system thinks that an energy attack is ongoing at time $t$, with $t = kT$, $k \in \mathbb{N}$.

Regardless of how to detect energy attacks, the inputs to the detection system at node $i$ may only be voltage $V_{n,i}(t)$, current $I_{n,i}(t)$, and capacitor charge~$C_i(t)$.
This is because the device has no information about ambient energy at node $i$, indicated as $V_{a,i}(t)$ in \figref{fig:attack}, \emph{before} its transformation into usable energy by the harvesting mechanism.
This also means that whenever an attack described as a function $a_i(V_{a,i}(t))$ occurs, the detection system \emph{cannot} observe the attack directly, but it only has access to \emph{its effects} as they manifest after the harvesting processing. 
Moreover, the need to reduce energy consumption rules out techniques that require sharing information among different nodes.

Of the requirements above, \textbf{R1} is common and many anomaly detection techniques for time series can fulfill that~\cite{chandola2009anomaly}.
Differently, \textbf{R2} excludes techniques that require the anomaly to stabilize in the long term or even to stop occurring before they can return an indication that it did happen.
 Most important, however, is \textbf{R3}.
It rules out anomaly detection methods that demand large memories~\cite{dietterich2000ensemble} or significant processing power~\cite{langley1992analysis}.
In contrast, it demands reducing energy consumption.
The ideal solution, and yet the most challenging to achieve would be  one that, in the face of intermittent executions, \emph{does not cause additional energy failures}, which may require persisting state on NVM and indirectly cause an immense overhead that \emph{would not be present otherwise}.

Based on this analysis, we base our solution on approximate intermittent computing~\cite{umesh2021survey,bambusi2022case} and develop an \emph{approximate support vector machine} (\approxSVM) for detecting energy attacks.
Approximate intermittent computing adapts approximate computing to intermittent executions, providing a knob to trade energy consumption for accuracy.
We elect to use support vector machines in the first place because of their \emph{extreme accuracy for binary classification}~\cite{boser1992training,pradhan2012support}, which is precisely our case, and their \emph{optimal trade-off between accuracy of classification and resource consumption}~\cite{mahdavinejad2018machine,david2021tensorflow,banbury2020benchmarking}.
 
 

\subsection{ASVM for Attack Detection}

An \approxSVM is a support vector machine trained as a regular SVM, but capable of using a subset of the signal features for inference, trading accuracy for resource consumption.


 
\fakepar{Training} We train the \approxSVM using a total of 52 features by sampling every $T$ the input signals $V_{n,i}(t)$, $I_{n,i}(t)$, ~$C_i(t)$, and combinations thereof.
Example features include statistical parameters like averages and standard variations, up to various maximum likelihood estimators.
The features are computed over different sliding windows over the past $wT, w \in \mathbb{N}$ time instants, which we tune based on the energy source.
For example, in case of solar radiation, we use different windows to account for the behavior of the energy source over the last 24 h and the last 10 min.
Training happens on a regular machine using the SVM library of the \code{ScyPy} package.

The energy for computing the features greatly depends on the input signals.
For example, many existing platforms for intermittent computing provide efficient ways to probe the onboard capacitor $C_{i}(t)$~\cite{flicker}, consuming negligible energy.
Differently, probing the energy harvester for the instantaneous supply voltage, as required to measure $V_{n,i}(t)$, requires a separate call to a dedicated ADC, with higher energy cost.
Based on the platform at hand, one may compute $I_{n,i}(t)$ from $V_{n,i}(t)$ without any further hardware interaction.


\fakepar{Inference} If we were to run the resulting model as a regular SVM, it may happen that running the energy attack detection step becomes the cause of an energy failure, prompting the system to dump the state on NVM to resume the work once energy is newly available.
The energy required for NVM operations is significant and even if it is not directly required for detecting energy attacks, its overhead is in fact indirectly caused by that process exceeding the available energy.
Even worse, if the energy failure happens in the middle of running the attack detection process, the output returned once the device resumes may not even reflect the current situation. 

To mitigate these occurrences, we profile the energy required for computing each of the 52 features, including the cost for hardware interaction, and analytically study the contribution to the overall accuracy each of them provides~\cite{epic}.
Next, we order the 52 features according to the ratio between their contribution to the resulting accuracy and their energy cost.
This means that features that come first are those with the lowest cost per unit of contributed accuracy. 

Whenever the detection system is to run, we probe the capacitor for the energy level and determine the first~$m$ features, out of the total 52 features, we can afford for classification \emph{without causing an energy failure}.
As probing the capacitor is required anyways, the energy cost for doing so is not considered in the energy cost profile of any of the features.
We then compute the classification incrementally and return the classification determined using $m \leq 52$ features.

By placing the call to the energy attack detection system as \emph{last} in an application period, we ensure that the attack detection process only runs with the energy ``leftovers'', without ever causing an energy failure that would not be there already in the original application.
Notably, it is possible to mathematically tie the number of features used for classification at run-time to the expected accuracy~\cite{bambusi2022case}, providing a framework to reason quantitatively on the tradeoff between energy cost and accuracy of detection.


\section{Evaluation}
\label{sec:eval}

Our evaluation is entirely based on real-world energy traces and accounts for 500K+ data points.
The analysis is three-pronged.
In \secref{sec:evalSynthetic}, we evaluate the accuracy and system performance of the \approxSVM by generating energy attacks in a synthetic manner, that is, by varying the statistical parameters representing how an attacker may manipulate the energy signal.
This is instrumental to assess the general behavior of our design in a multitude of conditions.
In \secref{sec:evalConcrete}, we concentrate on accuracy and system performance of our design against the three specific kinds of attacks we analyze in \secref{sec:attacks}.
We conclude the evaluation in \secref{sec:evalUnknown} by measuring the performance against attacks whose patterns are never seen before, testing our design's ability to tackle previously unseen scenarios.
We describe next the setup, the metrics and energy traces, and the baselines.

\subsection{Setting}
\label{sec:setup}
Because of the highly non-deterministic behavior of energy sources, achieving perfect reproducibility when evaluating energy-harvesting systems is extremely challenging using real devices~\cite{EKHO}.
We thus opt for system emulation over hardware-based experimentation, as it enables better control on experiment parameters and allows us to carefully reproduce executions and energy patterns across the \approxSVM and the baselines.
Still, in the specific case of \secref{sec:evalConcrete}, we manage \emph{to run experiments with a real device and energy harvester}.

\fakepar{Setup} We use the custom Siren MSP430 emulator~\cite{siren} we rely on in \secref{sec:attacks}.
To isolate the impact of the energy attack detection system, we use a single capacitor architecture.
We extend the tool with a model of 64 Kbyte of FRAM NVM next to a 2 Kbyte SRAM space, corresponding to the memory layout of the MSP430-FR5969, often employed in intermittent computing~\cite{mementos, Hibernus, ratchet, alpaca, DINO}.
We account for the energy consumption per clock cycle of various operating modes of the MSP430-FR5969~\cite{msp430fr5969}, such as regular computation, (non-)vo\-latile memory read/write operations, and I/O.

The application workload is the same as \secref{sec:reality} and we use HarvOS~\cite{HarvOS} for checkpointing.
We do not emulate sensors and radio, but the time and energy overhead for both are synthetically accounted for at the end of every application round.
Note that the specific application processing is orthogonal to the energy attack detection.
The application is instrumental here to quantify the relative overhead of the energy attack detection system on the application's original performance.
In our case, we insert a call to the energy attack detection system at the end of \emph{every} application round, thus generating the highest relative overhead.

\fakepar{Metrics and traces} We compute four key metrics.
\begin{glossy_enumerate}
\item The \emph{percentage of false negatives (positives)} measures the accuracy of the detection process as the fraction of energy attacks that are definitely missed (wrongly detected).
\item The \emph{time to detection} is the number of application rounds between attack injection and when the detection system signals the attack.
  This measures how rapidly a given technique realizes that the signal represents an anomaly\footnote{False negatives are excluded from aggregate statistics.}.
\item The relative \emph{energy overhead} is the additional energy consumption due to the energy attack detection system, compared with running the application alone.
\end{glossy_enumerate}

We feed the emulator with twelve diverse energy traces, offering a mixture of energy source, harvesting technology, and setting.
In the following, the terms ``indoor/outdoor'' coupled with ``static/mobile'' refer to the location and mobility of the harvester unit, respectively.

The M-RF trace is from Mementos~\cite{mementos} and is recorded using a WISP device and a Powercast transmitter~\cite{sample2008design}.
Four traces are from EPIC~\cite{epic}  and are recorded using a mono-crystalline solar cell in settings including outdoor mobile (E-SOM), indoor mobile (E-SIM), outdoor static (E-SOR), and indoor static (E-SIR).
The M-VIB and M-TEG traces are obtained using a ReVibe modelD kinetic energy harvester~\cite{revibe-modelD} and a Thermalforce 254-150-36 thermoelectric energy harvester~\cite{teg}, respectively.
The remaining traces are from the development of the Bonito protocol~\cite{geissdoerfer2022learning} and account for scenarios involving solar cells (B-JOG and B-OFI) and piezoelectric (B-STA, B-CAR, B-WAS) harvesters.

We use a 75/25 split between training and test data for every trace~\cite{mohri2018foundations}.  
During each experiment, we inject no attack for 0.5~h of simulated time to let estimators stabilize.
Experiments last between 12~h and 24~h of simulated time.

\fakepar{Baselines} We compare the performance of the \approxSVM in detecting energy attacks against an equivalent SVM that uses the same model with the regular inference step, called \fullSVM, as well as two additional baselines.

\revision{One baseline is the \emph{k-nearest neighbors algorithm} (\knn)~\cite{mohri2018foundations}, together with hyperparameter optimization to select $k$.
\knn is often used for anomaly detection of time series~\cite{chandola2009anomaly}.
As in existing work~\cite{chandola2009anomaly}, we obtain the classification by majority vote across the 52 features.}
The other baseline is called \emph{isolation forest} (\forest) and is known to provide accurate anomaly detection with a linear time complexity and very limited memory consumption~\cite{liu2008isolation}.
These features would be useful to meet requirement \textbf{R3} of \secref{sec:detection}.

\subsection{Synthetic Attacks}
\label{sec:evalSynthetic}

\begin{figure}[tb]
  \centering
  \vspace{-2mm}
  \subfigure[{False negatives  [\%].}]{
    \label{fig:falseNeg}
   \includegraphics[width=.9\linewidth]{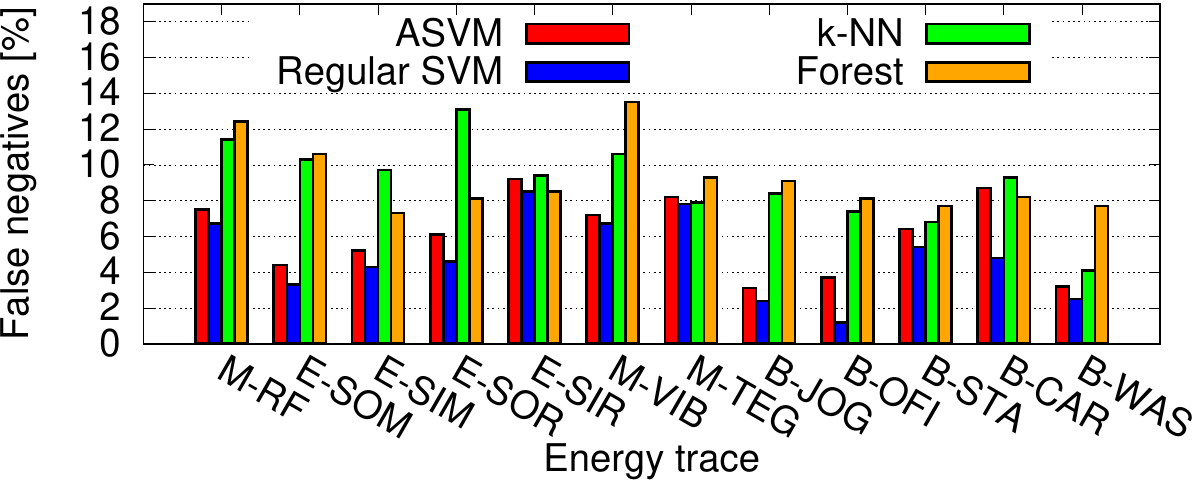}
  }
  \subfigure[{Energy overhead [\%].}]{
    \label{fig:overhead}
    \includegraphics[width=.9\linewidth]{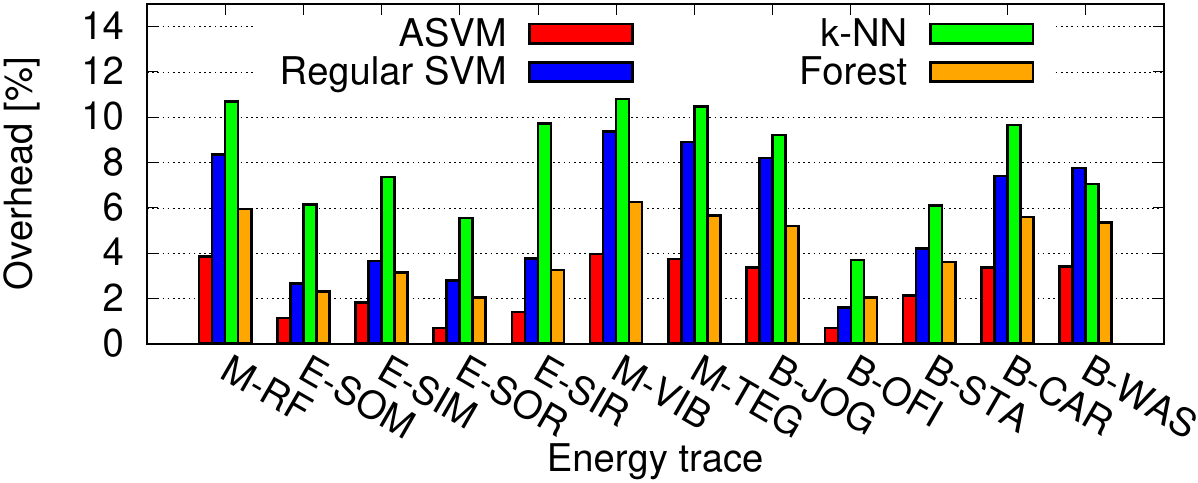}
 }\vspace{-2mm}
  \subfigure[{Time to detection  [periods].}]{
    \label{fig:time}
   \includegraphics[width=.9\linewidth]{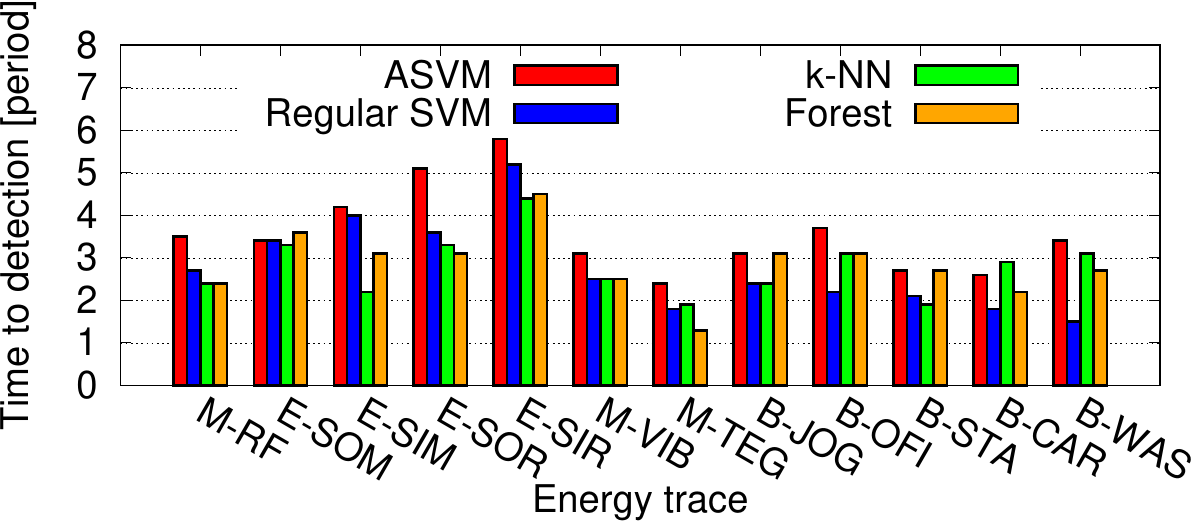}
  }
   \vspace{-3mm}
    \caption{Performance of \approxSVM, \fullSVM, \knn\ and \forest as a function of energy trace. \capt{False positives are limited in all designs. False negatives are much lower for both SVM-based solutions, with a slight loss of accuracy for \approxSVM. The latter comes in exchange of much lower energy overhead. Whenever a design does detect an attack, \knn and \forest tend to be slightly more rapid.}}
  \label{fig:resultTrace}
  \vspace{-3mm}
\end{figure}

\begin{figure*}[tb]
  \centering
  \subfigure[{False negatives as a function of $k$, $h$ is equal to one.  [\%].}]{
    \label{fig:falseNegK}
   \includegraphics[width=.45\linewidth]{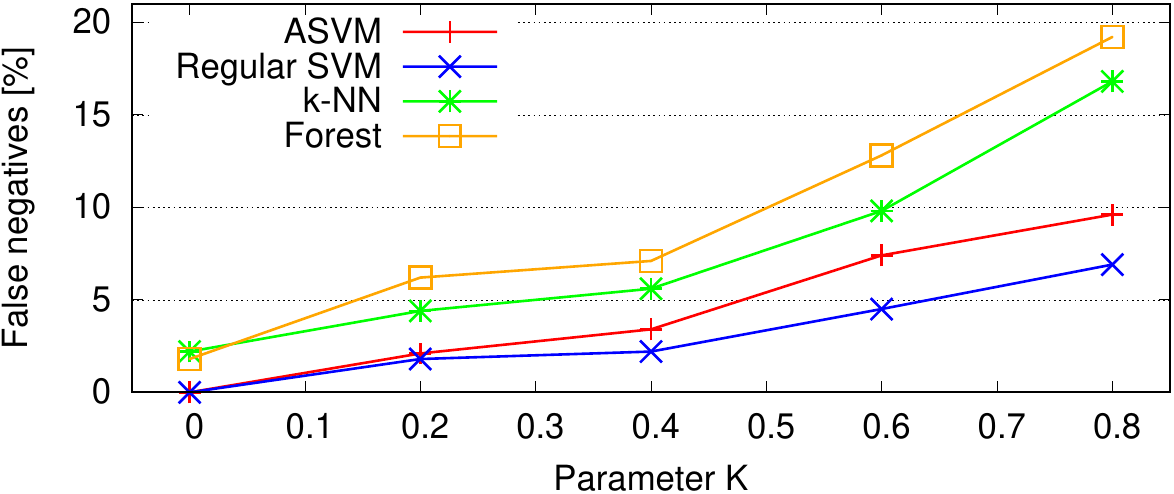}
  }
  \subfigure[{False negatives as a function of $h$, $k$ is equal to one. [\%].}]{
    \label{fig:falseNegH}
    \includegraphics[width=.45\linewidth]{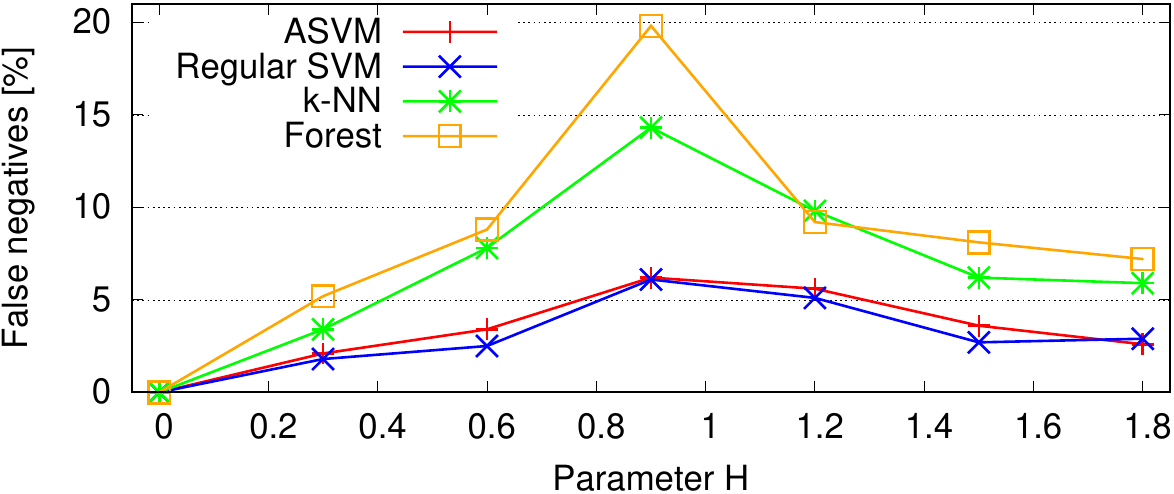}
  }
   \vspace{-3mm}
    \caption{Percentage of false negatives for \approxSVM, \fullSVM, \knn\ and \forest as a function of $k$ and $h$. \capt{The more the manipulated signal is similar to the original one, the more difficult is the detection and the accuracy decreases. Both SVM-based designs, however, consistently outperform both \knn\ and \forest, with the largest improvements corresponding to the cases with $k\approx1$ or $h\approx1$.}}
   \vspace{-3mm}
  \label{fig:resultTraceKH}
\end{figure*}

\revision{We create a generic setting that does not explicitly model any of the specific attacks in \secref{sec:attacks}.
  The results we present next are hence representative of our design's performance in the absence of specific information on the expected attacks.}

We inject energy attacks by considering the original trace as a signal $V_{n,i}(t)$ with mean $\mu$ and variance $\delta$ and by manipulating the latter two, changing the mean as $\mu' = k\mu$ and/or the variance as $\delta' = h\delta$, with $0 \leq k,h; k < 1; h< 2; h,k \in \mathbb{Q}$.
Intuitively, $h >1$ means increasing the randomness in the signal. 
Attack occurrence is drawn from a Poisson distribution with arrival rate $\lambda$ of one every 10 minutes; the duration is drawn from a Normal distribution with mean 3 min.
These parameters model the attacks of the exploratory study of \secref{sec:reality}.
The application round repeats every minute.


\fakepar{Results} \figref{fig:resultTrace} shows the results we obtain as a function of the energy trace. 
The percentage of false positives, which is not graphically reported for brevity, is limited across all designs and energy traces, never even reaching 1\% of the cases.
This is reassuring: whatever countermeasure the systems applies when detecting an attack, as further discussed in \secref{sec:defense}, the price to pay for that is rarely paid unnecessarily.

Different considerations apply to the results for false negatives, show in \figref{fig:falseNeg}, which represent a more critical situation: the system is under attack, but it is unaware of that.
The performance of both SVM-based solutions is markedly better than the other two baselines for most of the energy traces.
The only exceptions are the E-SOR and B-STA traces, obtained using solar cells in an indoor setting, where the performance is comparable among all designs.
The chances that temporary occlusions of the solar cells occur in this setting, for example, because of people passing by, is much higher than elsewhere.
It is thus difficult to separate these legitimate variations in energy supply from short-lived energy attacks.

Crucially, \figref{fig:falseNeg} demonstrates that the loss of accuracy, measured in terms of false negatives, due to using fewer features in \approxSVM is extremely limited compared to the \fullSVM.
In the worst case, shown by the B-OFI trace, the \approxSVM is only 2.4\% less accurate than the \fullSVM.
In return, the \approxSVM imposes a much lower energy overhead: \emph{between half and one third} of that of \fullSVM, as shown in \figref{fig:overhead}.
Two factors concur to this result: \emph{i)} the \approxSVM uses \emph{first} the features that contribute the most to an accurate result; and \emph{ii)} the \approxSVM is inherently adaptive: if sufficient energy is available, it uses all available features and the performance is the same as the \fullSVM.
As for the other baselines, their energy overhead tends to be higher than \approxSVM, with \forest generally outperforming \knn due to the linear time complexity.

\figref{fig:time} shows, on the other hand, the one weakness of both SVM-based approaches, that is, a slightly worse performance in time to detection compared to the other baselines.
On average across all energy traces, \approxSVM (\fullSVM) takes roughly 12\% (8\%) additional time to detect that an energy attack is occurring.
Given the absolute numbers at hand, however, this additional time rarely corresponds to more than one application period.
Provided the attack is eventually detected, this means the application spends limited time without knowing about that.

Given the accuracy and energy performance of the \approxSVM, the additional time to detect an attack is a fair price to pay, given that the other two baselines are slightly faster to detect an attack \emph{when they do detect one}, but often miss the detection altogether, as shown earlier in \figref{fig:falseNeg}.
The \approxSVM is slightly slower, but significantly more accurate, that is, ``better (slightly) late than never''.

\begin{figure}[tb]
  \centering
  \vspace{-2mm}
  \subfigure[{False negatives  [\%].}]{
    \label{fig:falseNegAttacks}
   \includegraphics[width=.9\linewidth]{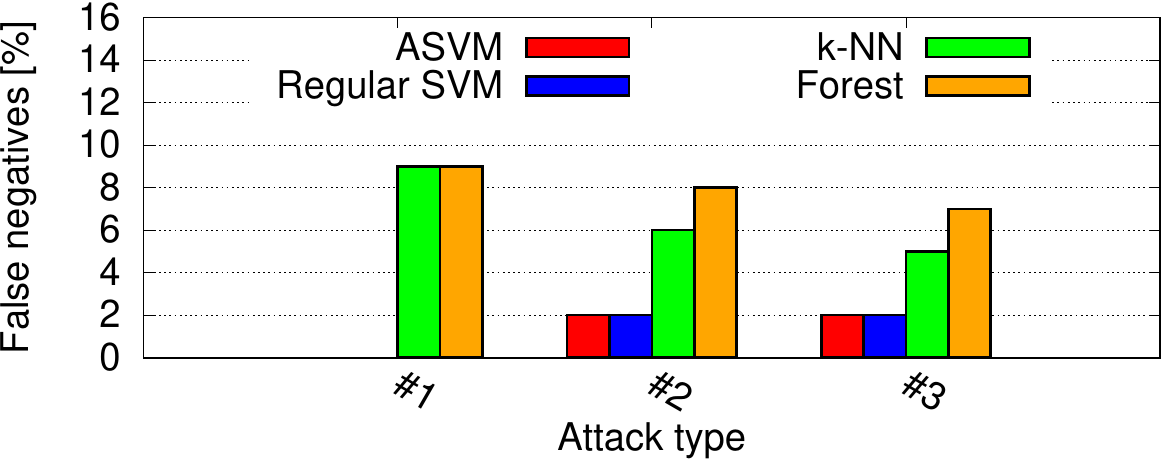}
  }
  \subfigure[{Energy overhead [\%].}]{
    \label{fig:overheadAttacks}
    \includegraphics[width=.9\linewidth]{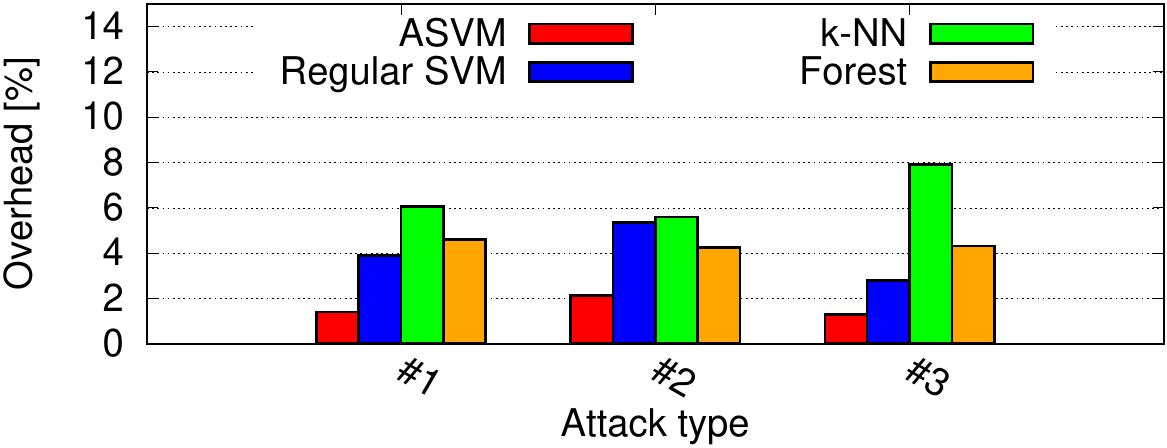}
 }\vspace{-2mm}
  \subfigure[{Time to detection  [periods].}]{
    \label{fig:timeAttacks}
   \includegraphics[width=.9\linewidth]{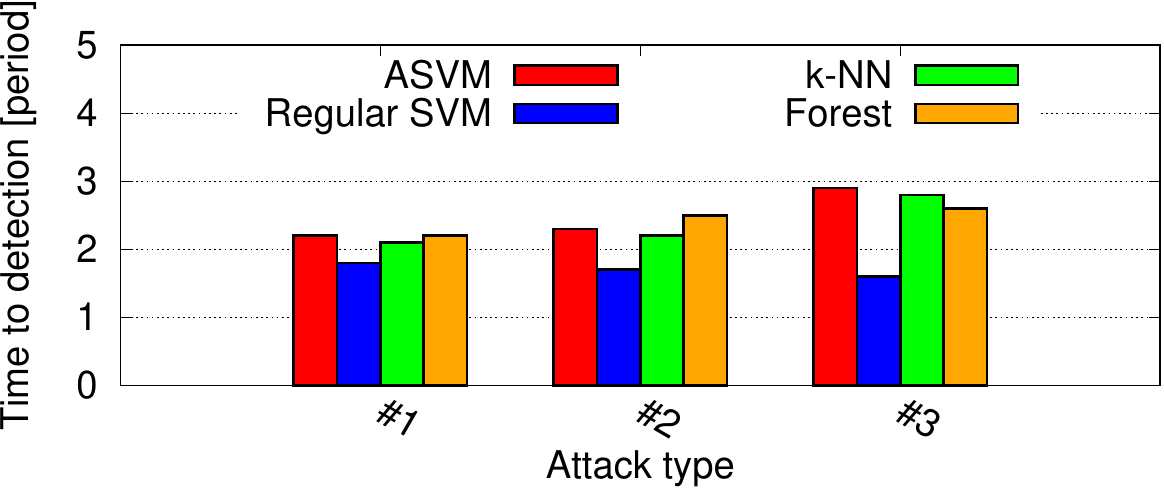}
  }
   \vspace{-3mm}
   \caption{Performance of \approxSVM, \fullSVM, \knn\ and \forest for the three attacks of \secref{sec:attacks}. \capt{False positives are still limited, with the \approxSVM returning no false positives at all.
The \approxSVM is as accurate as the \fullSVM but imposes a smaller energy overhead. Time to detection for the \approxSVM is similar to \knn and \forest, and only marginally higher than \fullSVM for attack \#2 and \#3.}}
   \vspace{-3mm}
  \label{fig:resultTraceAttacks}
\end{figure}

\figref{fig:resultTraceKH} provides a different view on the results, plotting the percentage of false negatives depending on $k$ and $h$.
We concentrate solely on the false negatives because the false positives are limited across the board, as discussed before, whereas energy overhead and time to detection are largely independent of $k$ and $h$.
\figref{fig:falseNegK} shows that both SVM-based designs consistently outperform the baselines regardless of $k$ and $h$.
The increasing trend is due to attacks with smaller $k$ being easier to detect: the smaller the $k$, the more different is the manipulated signal compared to the original one.
The corner case is with $k=0$, which corresponds to completely zeroing the energy signal, for example, modeling an occlusion of a solar cell, which all designs recognize accurately.
The more $k$ approaches 1, that is, the closer is the manipulated signal to the original one, the wider is the gap between the SVM-based designs and the baselines.

Similar considerations apply to \figref{fig:falseNegH}, plotting the percentage of false negatives as a function of $h$.
Again, the more $h$ approaches~1, which means the more the manipulated signal is similar to the original one, the less accurate is the detection.
The gap between the \approxSVM and either \knn or \forest is largest precisely around $h\approx1$: we obtain the best relative improvement in the most difficult setting. 

\subsection{Concrete Attacks}
\label{sec:evalConcrete}

We verify the performance of the \approxSVM against the three concrete attacks we analyze in \secref{sec:attacks}.
For attack \#1, we use the same \emph{configuration and real hardware} as in \secref{sec:attacks} and examine 22 instances of this attack.
For attack \#2 and \#3, we generate 100 instances each using the same setup and traces of \secref{sec:setup} as input.
The duration of each attack \#2 is uniformly chosen between 10 and 40 sec.

There is no guarantee that the attacks we generate are necessarily successful.
By monitoring the signal $V_n(t)$ during the experiments, we measure that 16 instances (out of 22) are successful for attack~\#1, 88 instances (out of 100) are successful for attack~\#2, and 38 instances (out of 100) are successful for attack~\#3.
This models a realistic setting where the attacker attempts to exploit a certain vulnerability, but does not always or consistently succeed.

\begin{figure*}[tb]
  \centering
  \subfigure[{False negatives as a function of $k$, $h$ is equal to one [\%].}]{
    \label{fig:falseNegKUnknown}
   \includegraphics[width=.45\linewidth]{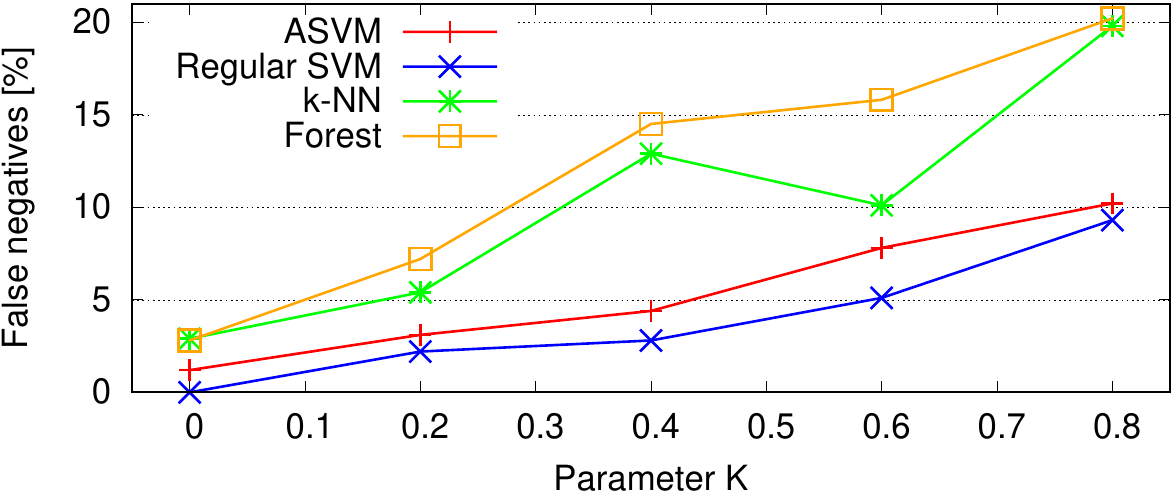}
  }
  \subfigure[{False negatives as a function of $h$, $k$ is equal to one [\%].}]{
    \label{fig:falseNegHUnknown}
    \includegraphics[width=.45\linewidth]{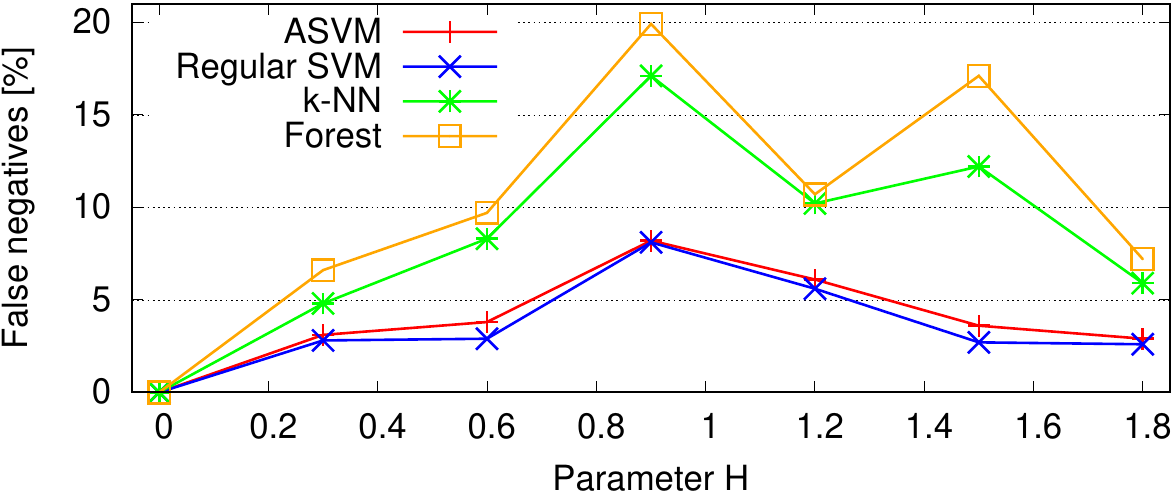}
  }
   \vspace{-3mm}
    \caption{Percentage of false negatives for \approxSVM, \fullSVM, \knn\ and \forest as a function of parameter $k$ and $h$, when using non-overlapping parameter settings for training data and test data. \capt{Unlike the other baselines, both SVM-based techniques are robust when tackling previously unseen problem instances.}}
  \label{fig:resultTraceKHUnknown}
\end{figure*}

\fakepar{Results} \figref{fig:resultTraceAttacks} shows the results.
The trends of \secref{sec:evalSynthetic} are confirmed. 
False positives, not shown for brevity, top to 3\% of the cases, and yet the \approxSVM, despite the approximate processing, returns no false positives at all.
Crucially, \figref{fig:falseNegAttacks} shows that the \approxSVM has the \emph{exact} same accuracy as the \fullSVM.
We conclude that the additional features processed by the \fullSVM are not sufficient to sway the classification compared to the \approxSVM, and hence the related processing entirely represents unnecessary overhead.  

The impact of the unnecessary overhead shows in \figref{fig:overheadAttacks}, where the \approxSVM outperforms all other designs, with \forest being the 2nd most efficient technique as seen in \figref{fig:overhead}.
Finally, \figref{fig:timeAttacks} demonstrates that, at least for these three specific attacks, the time to detection for the \approxSVM is comparable to \knn and \forest, and only the \fullSVM performs somehow better for attack~\#2 and~\#3, while being roughly as fast for attack~\#1.

\subsection{Unknown Attacks}
\label{sec:evalUnknown}

We evaluate what accuracy we may obtain against unknown problem instances.
We repeat the experiments of \secref{sec:evalSynthetic} with a different split between training data and test data.
We use as training set all problem instances where parameter $k$ or $h$ are such that $k \in \{0.2, 0.6\}$ and $h \in \{0.6, 1.2, 1.8\}$.
All other parameter setting for either $k$ or $h$ are exclusively used to generate test data.
This has two effects: \emph{i)} it reduces the size of the training data, and \emph{ii)} the designs we test are confronted with unseen patterns of the energy signal, which might be legitimate or represent attacks.  

\fakepar{Results} \figref{fig:resultTraceKHUnknown} shows the results.
For the reasons explained earlier, we focus on the percentage of false negatives as a measure of accuracy.
Compared to \figref{fig:resultTraceKH}, the absolute values are generally larger, likely because of the reduction in the size of the training data, yet the trends remain largely the same.
The observations we outline earlier, especially on the limited loss of accuracy of the \approxSVM compared to the \fullSVM, as well as on the better performance of both compared to the other baselines, remain.

Most importantly, the parameter settings that do \emph{not} appear in the training set represent no particular outlier in \figref{fig:resultTraceKHUnknown} for either of the SVM-based designs.
This is not the case for the other baselines, as seen in \figref{fig:falseNegKUnknown} when $k=0.4$ and in \figref{fig:falseNegHUnknown} when $h=1.5$.
This provides evidence of the general robustness of our design also when facing previously unseen problem instances.


\section{Inspiring Defense}
\label{sec:defense}

The system should defend against energy attacks and limit their negative effects.
Designing defense techniques is a manifold problem, whose implications possibly percolate through both software and hardware layers.
We discuss next key dimensions of the design space and offer some food for thoughts, inspiring follow-up work in the area.

\fakepar{Metrics} Application requirements dictate what metric is representative of the negative effect that must be mitigated.

Consider a scenario where \emph{energy} is the major concern, hence the ratio between energy consumed and useful work is to be minimized.
Defense techniques may be developed in this scenario by applying mixed-criticality concepts~\cite{burns2017survey}, that is, by splitting code functionality or application tasks in critical and non-critical ones.
During an attack, the latter may be suspended, thus shifting the reduced energy budget towards critical functionality.
In the case of attack~\#1, for example, this may yield the conditions for successfully completing checkpoints, resolving the livelock.

Say the amount of \emph{collected data} is crucial, regardless of how energy is spent.
One may employ concepts of context-oriented programming~\cite{hirschfeld2008context}, which also exist for low-power embedded systems~\cite{afanasov2014context}, to dynamically change the application behavior.
For instance, the system may temporarily log data locally instead of using the radio for wireless transmissions.
In the case of attack \#3, for example, doing so would lower the pressure on collision avoidance mechanisms, thus improving packet delivery ratio, while providing the additional benefit of preventing the attacker to gain information by sniffing wireless packets.

\fakepar{Attack-specific and general techniques} One may design techniques that mitigate particular negative effects.
These require an additional step to identify the type of attack.

Examples are techniques that mitigate attacks preventing a device from making progress, that is, akin to attack~\#1.
Recognizing this kind of attack may be achieved by instrumenting the code with intermittence-aware programming constructs~\cite{maioli2020intermittence}.
Programmers can use these constructs to count how often a certain instruction in the program is executed together with the same system state.
By properly setting a threshold based on the number of transient livelocks that are normally expected based on environment dynamics, one may recognize that a livelock is occurring when it is not expected.
Adapting techniques that dynamically adjust the activation threshold, as in the Hibernus++ checkpointing system~\cite{Hibernus++}, may provide a way to resolve the livelock. 

One may counteract attacks exploiting the task structure of programs, as in attack~\#2, by adapting task splitting and coalescing techniques~\cite{coala,ink}.
These allow to dynamically change the task structure of programs, which is normally static, by dividing or merging subsequent tasks.
This provides an additional degree of freedom to handle temporary situations where producer tasks run much more frequently than consumer ones.
Mixed-criticality concepts~\cite{burns2017survey} may also help: as an example, temporarily suspending tasks that only perform logging of system functionality may make room for sensed data in the local data buffers, hence postponing the time when data starts overflowing.

On the other hand, we argue that generally-applicable defense techniques may take inspiration from energy management in mobile devices~\cite{hoque2015modeling}.
Mobile operating systems feature sophisticated techniques to handle situations of energy scarcity, especially when the battery is about to run out.
These include tuning a number of hardware knobs, which are normally not available on resource-constrained embedded platforms, as well as software techniques such as throttling the execution of a subset of system processes.
The latter effectively represent a generalization of many of the attack-specific techniques we discuss.
These techniques may be adapted to intermittently-computing devices.

\fakepar{Hardware and software defense} The defense techniques outlined above are mainly implemented in software.

A natural option at the hardware level is to rely on multiple energy sources.
Existing hardware platforms~\cite{flicker} and deployed battery-less IoT systems~\cite{water-deployment-microbial-fuel-cell,tethys,soil-termoelectric,sensys20deployment} seldom rely on multiple energy sources, yet prototypes exist~\cite{liu2021hybrid}.
As we discuss next, the feasibility of energy attacks is strictly tied with the nature of the energy source.
Combining energy-rich sources that may be attack vectors, with energy-poor sources that are exceedingly difficult to employ as attack vectors, may provide an effective combination to sustain long-term operation in regular conditions, while mitigating the effects of energy attacks when they occur.


\section{Discussion}
\label{sec:discussion}

The feasibility of energy attacks depends not just on the abilities of the attacker, but also on the energy source.

Certain energy sources are simple to control, such as RF energy: Powercast receivers and transmitters can be hooked to a regular machine and controlled programmatically.
Other sources may be controlled under given conditions.
Light, for example, is controllable in indoor environments.
An attacker may gain control of the lighting infrastructure in a building, as it is often assumed in visible light communications~\cite{jovicic2013visible}, and use that to convey the attack.
This setup should, however, possibly work in combination with other sources of light that cannot be as easily controlled, for example, solar radiation entering a building from the outside.
At the other extreme, energy sources exist where the physical medium cannot change sufficiently rapidly, for example, temperature gradients.
Using these sources as attack vectors is likely to be exceedingly difficult.

An orthogonal dimension relates to time.
Depending on the source, the attacker may need a variable number of attempts before an attack succeeds.
For instance, how rapidly attack \#1 using RF energy is going to succeed depends on the choice of $\delta$, which drives the search of the phase corresponding to the opposing signal.
If $\delta$ is too large, the procedure may never yield the conditions for the attack to succeed; the attacker then chooses a smaller $\delta$ and repeats the procedure.
A similar consideration applies to attack \#2 and $\Delta t$.

Lacking source code information may make setting up certain energy attacks extremely laborious, whereas the same information is not as fundamental in other cases.
For attack~\#3, for example, it would suffice to know that the application logic is the same across all devices and unfolds as the usual sense-process-transmit loop. 


\section{Conclusion}
\label{sec:end}

We studied how exerting limited control on ambient energy provisioning to battery-less IoT devices may be used as an attack vector.
We provided experimental evidence of three types of energy attack and discussed the corresponding software vulnerabilities.
Next, we designed a way to detect energy attacks that is accurate, timely, and imposes a limited energy overhead, enabling detection right on the resource-constrained IoT devices.
Our evaluation, entirely based on real-world energy traces, shows that the \approxSVM detects energy attacks with 92\%+ accuracy, that is, up to 37\% better than the baselines, and with up to one fifth of their overhead.


%
%
%

\bibliographystyle{IEEEtran}
\bibliography{biblio,bibliography}

\end{document}